\documentstyle[preprint,aps]{revtex}

\renewcommand{\Im}{\mathop{\rm Im}\nolimits}

\begin{document}
\draft
\title{Conductance and conductance fluctuations of \\
       narrow disordered quantum wires}

\author{K Nikoli\'c and A MacKinnon}
\address{The Blackett Laboratory, Imperial College, 
           Prince Consort Road, London SW7 2BZ, UK}

\date{\today}
\maketitle

\begin{abstract}
In this paper we present and discuss our results for the conductance
and conductance fluctuations of narrow quantum wires with two types of
disorder:  boundary roughness (hard wall confining potential) and
islands of strongly scattering impurities within the bulk of the wire.
We use a tight--binding Hamiltonian to describe the quantum wire,
infinite perfect leads, a two--terminal Landauer--type formula for the
conductance, and the recursive single-particle Green's function
technique.  We find that conductance quantization is easily destroyed
by strong scattering. We also find that Anderson localization poses
a serious restriction on the high carrier mobility predicted in quantum
wires.  Conductance fluctuations in narrow quantum wires are not, in
general, universal (as in the metallic regime), but can be independent of
the wire length over a short range of lengths.

\end{abstract}

\pacs{73.20.Dx, 71.20.--b}

\narrowtext

 \newcommand{\etal}{{\it et al.\ }}
 \newcommand{\ie}{{\it i.e.\ }}
 \newcommand{\eg}{{\it e.g.\ }}
 \newcommand{\rms}{{\rm rms}}
 \newcommand{\Tr}{\mathop{\rm Tr}\nolimits}
 \newcommand{\lav}{\langle}
 \newcommand{\rav}{\rangle}

 \newcommand{\pr}[1]{{\it Phys.\ Rev.} {\bf #1}}
 \newcommand{\jpcm}[1]{{\it J.\ Phys.:\ Condens.\ Matter} {\bf #1}}
 \newcommand{\jpc}[1]{{\it J.\ Phys.\ C: Solid State Physics} {\bf #1}}
 \newcommand{\ssc}[1]{{\it Solid\ State\ Commun.} {\bf #1}}
 \newcommand{\epl}[1]{{\it Europhys.\ Lett.} {\bf #1}}
 \newcommand{\zpb}[1]{{\it Z.\ Phys.\ B} {\bf #1}}
 \newcommand{\prs}[1]{{\it Proc.\ Roy.\ Soc.\ Lond.} {\bf #1}}
 \newcommand{\pps}[1]{{\it Proc.\ Phys.\ Soc.} {\bf #1}}
 \newcommand{\ibmj}[1]{{\it IBM\ J.\ Res.\ Develop.} {\bf #1}}

\section{INTRODUCTION}
\label{sec:intro}

Many novel transport phenomena have been revealed in mesoscopic
low-dimensional structures. The considerable interest in submicron
electronic structures has been motivated by the expectation that
potentially useful new devices could be invented.  Very sophisticated
techniques have been developed, such as the various types of epitaxial
growth, lithography, ion implantation, etching and cleaving,
\mbox{etc.} in order to make these small size structures.  However,
none of them can produce perfect quantum wires. For example, GaAs/AlAs
quantum well wires grown on a vicinal surface using molecular--beam
epitaxy (MBE) \cite{petroff84} have two characteristic types of
disorder. The interface between the GaAs and the AlAs regions is not
smooth and, in addition, within the region of nominally pure GaAs,
there will be islands of AlAs. The question of how compositional
disorder affects the transport properties of quantum wires is
important.

The conductance of narrow ballistic channels, or  quantum point
contacts in a 2D electron gas (2DEG) is quantized in integer multiples
of $2e^2/h$ \cite{vanwees88}.
 However, this simple step--like form for the conductance as a function
 of the Fermi energy, occurs when the transition between the wide leads
and the narrow channel is adiabatic\cite{glazman88}.  Nonadiabatic, \ie
mode-mixing transport through the constriction produces some additional
features in the conductance diagram
\cite{kirczenow88,szafer89,PhDThesis}.  The conditions for adiabatic
transport are readily achieved in the experiments. Disorder, however, can
quite easily destroy conductance quantization.  Poor quantization is
believed to be mainly caused by backscattering.  Backscattering, at low
temperatures, is produced by the impurities within the wire and/or by
the rough wire edge. It can also be caused indirectly through resonant
states trapped in the wire, which may be created by the random field of
impurities outside the wire \cite{nixon91,laughton91}.  However,
forward scattering does not generally harm the conductance
quantization, if all conducting modes are fully occupied
\cite{laughton91}.  Small--angle backscattering predominates in the
highest subbands and is usually considered responsible for the
destruction of the conductance quantization. When only the lowest 1D
subband is occupied, however, the number of states into which an
electron can be scattered by disorder is reduced. This gave rise to the
prediction of a large electron mobility in a confined electron system
\cite{sakaki80}.  However, the conductance is very strongly influenced
by the quantum interference effects which start to emerge as elastic
scattering is  introduced in the ballistic regime.  In the  quasi--1D
case there is a much higher probability of multiple scattering from the
same site compared to the 2D or 3D case. This could cause Anderson
localization to become dominant.

The conductance fluctuations are expected to show peculiar behaviour in
narrow quantum wires \cite{tamura91,ando92}. In the metallic regime all
the dimensions of a sample are much larger than the mean free path $l$ and
the electron motion is well defined in all directions. Hence, the
perturbation theory approach based on Feynman diagrams, which yields
universal conductance fluctuations (UCF), is valid. However, the
transverse quantization in quantum wires  gives well resolved 1D
subbands if the width of the wire ($w$) is smaller than the mean free
path ($l$) and comparable to the Fermi wavelength $\lambda_F$.
Therefore, in general, UCFs are not expected in quantum wires with weak
disorder. On the other hand we know that in the case of strong disorder
when the wire length $L$ becomes greater than the localization length
$\lambda$, subband mixing is very strong and any quasi--1D feature of a
perfect system is virtually destroyed \cite{taylor91,nikolic93}. Then
it becomes essential to take localization effects into account.
Again, application of a perturbational method \cite{stone92}) is
inappropriate.

 Here we examine the influence of rough boundaries
(Section~\ref{conduc-rough}), impurities
(Section~\ref{conduc-islands}), and both types of disorder combined
(Section~\ref{conduc-real}) on the conductance and conductance
fluctuations of quantum wires at zero temperature and zero magnetic
field. All calculations were done using a Landauer--type formula for
the conductance.


\section{Model and Methods}

Our calculations of the conductance, described below, require detailed
structural information about the quantum wire. A direct way of
describing such a system would be by providing detailed geometric and
structural information about an actual wire. Such data can be represented in
a convenient form as a set of probability distributions and correlation
functions of some basic parameters of the wire (\eg the width, width
fluctuations, confining potential etc). The detailed structure can be
recovered, in a statistical sense, by generating wires using an
appropriate algorithm and the set of probability distributions and
correlation functions as an input.
 In order to examine the electronic properties of some realistic
 structures we have used structural information obtained in the
Monte-Carlo simulation of vicinal surface grown quantum well wires
by Hugill \etal \cite{hugill}.

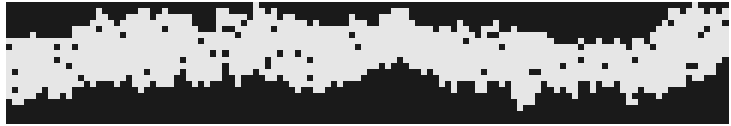
\begin{figure}
\caption{Plot of a section of the generated (real) quantum wire of average
width 10, with island concentration p=0.05.}
\label{wire}
\end{figure}

Quantum wires directly grown by epitaxial growth of a heterostructure,
usually (Al,Ga)As, by using more or less controlled generation of
terraces and steps (or corrugations) on semiconductor surfaces, seemed
very attractive \cite{petroff84,fukui87,miller92,notzel91}.
 This process was at one time considered very promising for the
eventual realization of very narrow (about a few nanometers)
quantum wires \cite{example}.  More recently attention has shifted to other
possibilities, such as V--grooves, but as the basic principles governing the
electronic structure are common to different sorts of wires we shall
concentrate here on MBE grown wires.
The kinetics of MBE can be successfully simulated on a computer
\cite{hugill,joyce}, provided that the values of the model parameters are
correctly estimated from the experimental data. This enables one to perform
Monte Carlo simulations of these wire structures and therefore, to
define structural disorder in the system \cite{taylor91,nikolic93}.
A section of a generated monolayer wire with an average width of 10 lattice
sites is shown in Fig.~\ref{wire}.  The effects of the various types of
compositional disorder considered here have implications for the
electronic behaviour of quantum wires fabricated by other techniques.

For the purpose of transport calculations the quantum wire is sandwiched
between two perfect leads. The same model using a tight--binding,
nearest--neighbour Hamiltonian is used  to describe both the quantum wire and
the leads:
\begin{equation}
{\bf H}= \sum_{i} |i\rangle \varepsilon_i \langle i| +
\sum_{i,j \atop (i \ne j)} |i\rangle V_{ij} \langle j|
\label{eq:hamilt}
\end{equation}
where $|i\rangle$ is the localized `Wannier' state or atomic orbital on site
$i$, $ \varepsilon_i $ is the `site energy' and $V_{ij}$ is the hopping
matrix element between sites $i$ and $j$. We shall assume that $V_{ij}$ is
zero unless the $i$ and $j$ sites are nearest neighbours, when
$V_{ij} \equiv V$ (i.e. $V$ defines our unit of energy and the effective mass).

We define our lead--sample--lead system to lie along the $x$--axis. It can be
divided into slices along that direction, each of which   has $M$ sites (\ie a
cross-section
of the quantum wire).
The elastic scattering in the quantum wire  (which extends from
slice 1 to slice $L$)
is described by transmission probabilities $T_{mn}=|t_{mn}|^2$, which describe
the probability that an electron incident in channel (state) $n$ on the left
emerges
in channel $m$ on the right. The amplitude transmission coefficients $t_{mn}$
can be calculated by various means.  Here the formulation due to Ando
\cite{ando91} is used
\begin{equation}
t_{mn} = \sqrt{\frac{v_m}{v_n}} \left[ {\bf U}^{-1}(+)V{\bf G}_{0,L+1}
         [{\bf F}^{-1}(+) - {\bf F}^{-1}(-)] {\bf U}(+) \right]_{mn}
\label{eq:ando}
\end{equation}
where
\begin{equation}
{\bf F}(\pm)  =  {\bf U}(\pm){\Lambda}(\pm){\bf U}^{-1}(\pm),
\label{eq:f}
\end{equation}
and $v_n$ is the longitudinal velocity in subband $n$.
${\bf G}_{0,L+1}$ is the Green's function which couples the $0$th and the
$(L+1)$st
slice in our system (\ie the last slice in the left--lead and the first slice
in
the right--lead).
 The matrices
\begin{equation}
{\bf U}(\pm)=\left[{\bf u}_1(\pm) \cdots {\bf u}_M(\pm) \right]~,
\label{eq:Upm}
\end{equation}
and
\[ \Lambda (\pm) = \left( \begin{array}{cccc}
 \zeta_1(\pm) &   &   &  \\
   & \zeta_2(\pm) &   &  \\
   &  &        \cdots   &  \\
   &  &  &  \zeta_M(\pm)
\end{array} \right)~. \]
contain the eigenvectors and  eigenvalues respectively of the  eigenvalue
problem
\begin{eqnarray}
\zeta  \left( \begin{array}{l}
   {\bf C}_{J}   \\
   {\bf C}_{J-1}
\end{array} \right) = \left( \begin{array}{cr}
    V^{-1}\left( E{\bf I}-{\bf H}_0^{(1)} \right)  &  -{\bf I}     \\
          {\bf I}                                 &   {\bf 0}
 \end{array} \right)  \left( \begin{array}{l}
   {\bf C}_{J}   \\
   {\bf C}_{J-1}
\end{array} \right)
\label{eq:eigenproblem}
\end{eqnarray}
The perfect leads extend to
$-\infty$ and $+\infty$ along $x$-axis. In these asymptotic regions
the incident and transmitted states obey the Schr\"{o}dinger equation
\begin{equation}
(E{\bf I} - {\bf H}_J^{(1)}){\bf C}_J-{\bf V}{\bf C}_{J-1}-
{\bf V}{\bf C}_{J+1}=0
\label{eq:discr_schr}
\end{equation}
where ${\bf H}_J^{(1)}$ is replaced by the Hamiltonian of an isolated
ordered slice ${\bf H}_0^{(1)}$.
${\bf C}_J$ is a vector describing the amplitudes of the wavefunction on
the $J$th slice.
The superscript designates the length of the system.
Although ${\bf V}$ is generally a diagonal matrix for the nearest--neighbour,
simple cubic
model, in the case of purely diagonal disorder and zero magnetic field it
reduces to a scalar.
Due to translational invariance along the
$x$--axis, the solutions of Eq.~(\ref{eq:discr_schr}) for the perfect leads,
must be in
the Bloch form, \ie:
\begin{equation}
\zeta {\bf C}_{J-1}={\bf C}_J~,
\label{eq:c2}
\end{equation}
where $\zeta=\exp({\rm i}ka)$ and $a$ is the lattice constant.
The eigenvalue problem, Eq.~(\ref{eq:eigenproblem}), is a
combination of the Schr\"{o}dinger equation~(\ref{eq:discr_schr}) and
Eq.~(\ref{eq:c2}). The 2M eigenvalues ($\zeta$) and eigenvectors
(${\bf u}$) can be separated into two groups: left--going, $\zeta(-)$
and ${\bf u}(-)$, and right--going waves, $\zeta(+)$ and ${\bf u}(+)$.
If $\zeta < 1$, then from Eq.~(\ref{eq:c2}),  the solution is exponentially
decaying in the positive $x$--direction and describes right--going evanescent
modes.  The $\zeta > 1$ solutions describe
left--going evanescent modes. If $\zeta$ is a complex number then
the classification is done according to the sign of the matrix element of the
current density operator ($j$)
(see Appendix B in \cite{baranger89} ):
\begin{eqnarray}
j & = & \frac{e}{2{\rm i}\hbar}\left[
{\bf C}_J^{\dag}{\bf C}_{J+1} - {\bf C}_J{\bf C}_{J+1}^{\dag} +
{\bf C}_{J-1}^{\dag}{\bf C}_{J} - {\bf C}_{J-1}{\bf C}_{J}^{\dag} \right]
\nonumber \\
 & = & \frac{2e}{\hbar} |{\bf C}_J|^2~\Im\zeta
\label{eq:j}
\end{eqnarray}
since $|\zeta|=1$.
If $\Im\zeta > 0$ then $j>0$ and the wave is propagating to the right, and if
$\Im\zeta < 0$ then it is propagating to the left.

The Green's function ${\bf G}_{0,L+1}$ is calculated by using the recursive
method \cite{mackinnonZ85}:
\begin{eqnarray}
{\bf G}_{1,N+1}^{(N+1)} &=& {\bf G}_{1,N}^{(N)}{\bf V}_{N,N+1}
{\bf G}_{N+1,N+1}^{(N+1)}, \label{eq:G1L} \\
{\bf G}_{N+1,N+1}^{(N+1)} &=& \left[ E{\bf I} - {\bf H}_{N+1}^{(1)} -
{\bf V}_{N,N+1}^{\dag}{\bf G}_{N,N}^{(N)}{\bf V}_{N,N+1} \right]^{-1}\;.
 \label{eq:GLL}
\end{eqnarray}
Iterative calculations are performed by successively adding slices to the
end of the bar. This numerical technique has proved very reliable for the
Anderson localization problem \cite{mackinnonprl81,soukoulis82}. The initial
conditions reflect the environment into which the wire sample is embedded. The
first
slice of the quantum wire (slice 1) is coupled to the end
(slice 0) of the left hand lead, \ie to a semi-infinite perfect wire. So the
initial
condition for calculating (${\bf G}_{NN}$ in
Eq.~(\ref{eq:GLL})) is given by the diagonal block of the Green's function
(${\bf G}_d^{(\infty/2)}$)
at the end of a perfect bar that extends from $-\infty$ to 0 (see
Ref.~\cite{mackinnonZ85}):
\begin{equation}
{\bf G}_{00} =
{\bf G}_d^{(\infty/2)}(\mbox{L--lead}) =
  {\bf U}(+)\Lambda (+){\bf U}^{-1}(+)V^{-1}
\label{eq:G00}
\end{equation}
Similarly for the right hand lead:
\begin{equation}
{\bf G}_d^{(\infty/2)}(\mbox{R--lead}) =
  {\bf U}(+)\Lambda (-){\bf U}^{-1}(+)V^{-1} = {\bf S}_{\infty}^{-1}
\label{eq:S}
\end{equation}
where ${\bf S}_{\infty}^{-1}$ is the self-energy matrix,
which helps us to couple the right hand lead to the other end of conductor.
The effect of adding the whole right hand lead can be represented by the
Hamiltonian:
\[ \tilde{\bf H}={\bf H}_0+{\bf S}_{\infty}~ \]
in the final iteration of Eq.~(\ref{eq:GLL}). Iterations of
Eq.~(\ref{eq:G1L}) for ${\bf G}_{1N}$ begin with the unit matrix.

The formulation  (\ref{eq:ando}) can be further simplified.
If ${\bf F}(\pm)$ is substituted by Eq.~(\ref{eq:f})
and since the same eigenproblem (\ref{eq:eigenproblem}) describes both
left-going and right-going solutions, we get our final result:
\begin{equation}
t_{mn} = \sqrt{\frac{v_m}{v_n}}~\left[~{\bf U}^{-1}(+)V{\bf G}_{0,L+1}
 {\bf U}(+) [\Lambda (-) - \Lambda (+)]~\right]_{mn}~~.
\label{eq:my}
\end{equation}
Note that the result is not affected by the normalization of ${\bf U}(\pm)$.
This formulation for $t_{mn}$ easily yields transmission probabilities for
the case of a perfect wire of length $L$ between two perfect leads of
the same cross--section:
\begin{equation}
T_{mn} = |t_{mn}|^{2} = |\zeta_{m}(+)|^{2L} \delta_{mn}
\label{eq:Tperf}
\end{equation}

The conductance $G$, given by the two--terminal Landauer formula
\cite{landauer57,fisher81}, for spin--degenerate states is:
\begin{equation}
G= 2\frac{e^2}{h} \sum_{n=1}^{N_L} \sum_{m=1}^{N_R}|t_{mn}|^2~~.
\label{eq:conductance}
\end{equation}
The summations run over the open channels, of which there are  $N_L$ in the
left lead, and $N_R$ in the right lead.

The conductance fluctuations are quantified by the square root of the variance
\begin{equation}
 \rms (G) = \left( \langle G^2\rangle - \langle G\rangle^2 \right)^{1/2}
\label{eq:fluctuation}
\end{equation}
where $\langle\rangle$ denotes averaging over an ensemble of
samples,  with different realizations of disorder. In our calculations all the
quantum wires have a hard wall confining potential. Also for the site energy
of islands $\epsilon_{\mbox{\scriptsize{isl}}}\rightarrow \infty$ is assumed.
The temperature of
the system is always $T=0K$.

\section{The Influence of Boundary Roughness} \label{conduc-rough}
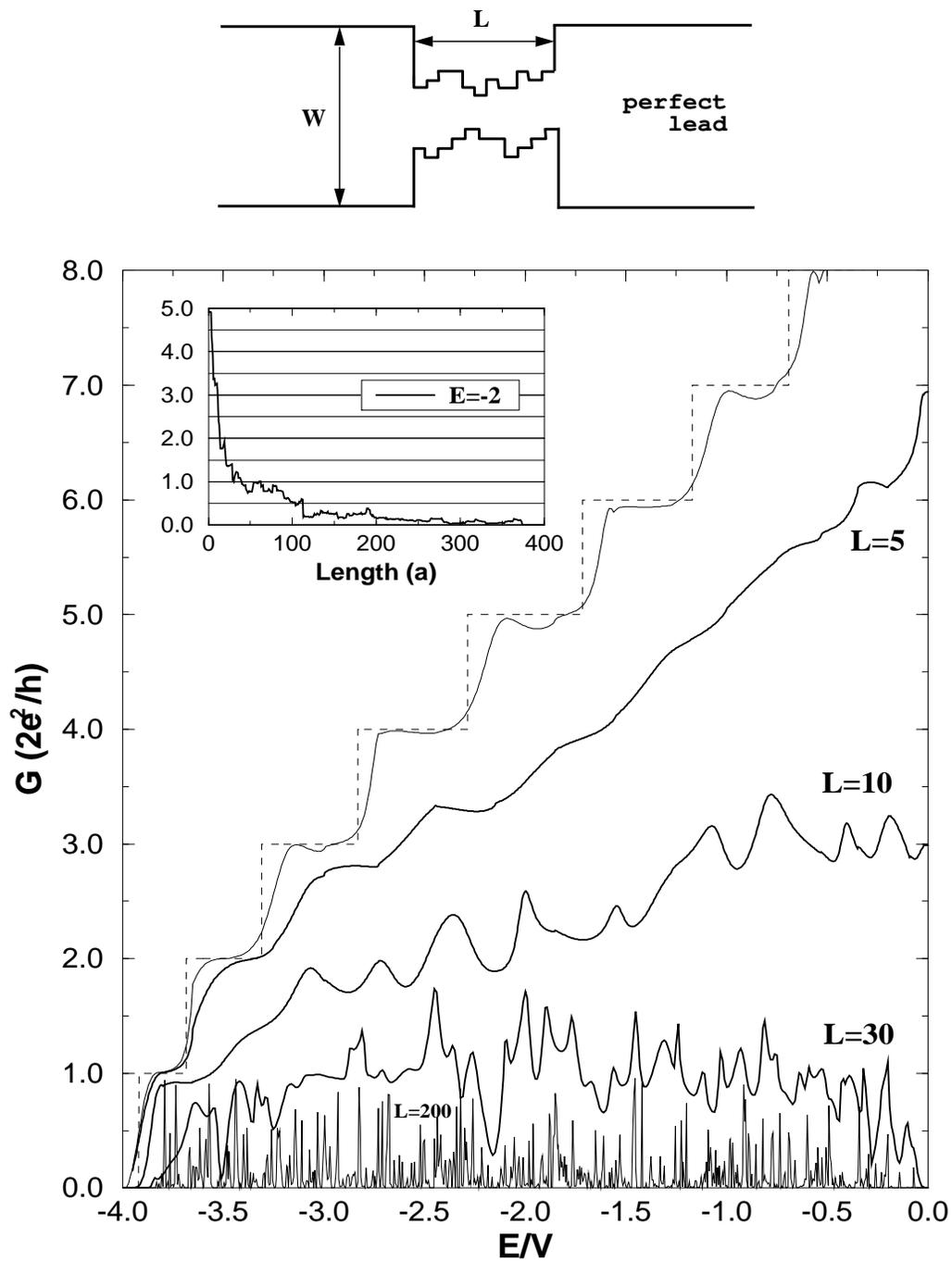
 \begin{figure}
 \caption{$G$ as a function of energy of a single sample of quantum wire
with rough edges and no island disorder, for wire lengths: $L=5, 10, 30$
and $200$, and average wire width $\lav w\rav=10$ and leads width $W=20$.
Reference step functions are the conductances of the perfect wire of the
width $w=10$, length $L=5$, and leads with $W=20$ (full line), and $W=10$
(\ie no difference between leads and wire -- broken line). The
inset is the conductance of a quantum wire
sample as a function of the wire length. The top picture shows the
geometry of the lead-wire-lead system used in the calculations.}
\label{rEone}
 \end{figure}

For the investigation of the transport properties of quantum wires with
rough edges we use the geometry of the system which is shown
at the top of Fig.~\ref{rEone}.
The average value of the width is taken to be $\langle w \rangle =10$ and the
width of the leads will be fixed at $W=20$ throughout this section.
Here we shall analyse results for four types of calculation:
conductance of a single quantum wire and ensemble average quantities related
to the conductance, as functions of Fermi energy $E$ and wire length $L$.
Firstly, we discuss the results for the conductance of a single quantum wire
sample,  presented in Fig.~\ref{rEone}.

The edge roughness destroys the conductance quantization steps, firstly near
the band
center for very small disorder (see case $L=5$ in Fig.~\ref{rEone}). The
deterioration of the quantization spreads towards the band edge both as the
disorder and as the length of the wire is increased.  In the mesoscopic regime,
when $l<L<\lambda$, the conductance
curve shows sample specific fluctuations as a function of energy and length.
However, the
amplitude of these fluctuations is of order $e^2/h$, independent of energy
(Fig.~\ref{rEone},
L=10, L=30) or length (see inset of Fig.~\ref{rEone}). This is a quantum
interference effect similar to the universal conductance fluctuations observed
in the mesoscopic regime \cite{lee85}) of other systems.
The particular value for the conductance is determined by the electron
wavelength (\ie electron energy), the length of the quantum wire and the actual
realization of the disorder in a
sample. The conditions for the destructive or constructive interference of an
electron wavefront are, generally, very sensitive to all of these parameters.
The scale of the sensitivity to energy, however, differs with the length of the
wire (which is evident from Fig.~\ref{rEone} for $L=10$ and $L=30$) or with
the level of disorder. This observation supports the idea that the
fluctuations do not arise from classical scattering from the rough boundary
but from the phase modulation of the electron wavefunction due to multiple
elastic
scattering in the wire \cite{takagaki92}. We estimate that the typical
spacing between peaks and valleys in the conductance as a function of energy
depends on the wire length as: $E_c\sim 1/L$. This is a weaker dependence on
the
wire length than in the case of  the universal conductance fluctuations in the
metallic regime, where $E_c\sim 1/L^2$ \cite{lee85}).  It should also be noted
that the conductance falls faster near the band center than elsewhere (this
will be clear from
the results for the average conductance). When the system is in the strong
localization regime, the conductance is reduced to a set of peaks
of different amplitudes, with maximum value of $2e^2/h$ (see Fig.~\ref{rEone}
$L=200$). The mechanism of electron transport is resonant tunnelling through
the quantum wire \cite{bryant91}. When the energy of an electron coincides with
an eigenenergy of the wire the electron can be transmitted through the wire via
this localized state. The height of the peak of $G$ depends on the overlap
between the wavefunctions of this spatially
localized state inside the wire and the propagating states in the leads. As the
length of the wire is further increased, the number of peaks and their
amplitudes reduce and their positions change.  The highest such peaks are
probably due to tunnelling through multiple resonant states, so--called
`necklace' states \cite{Pendry}.

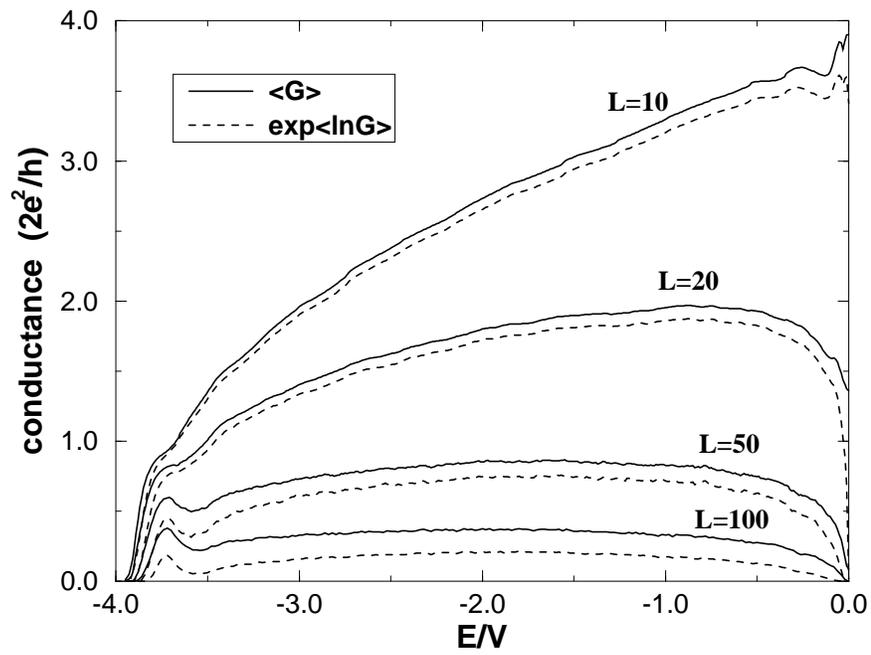
 \begin{figure}
 \caption{Average conductance as a function of energy, for quantum wires with
rough boundary and no island disorder, for different wire lengths:
$L=10$, $L=20$, $L=50$ and $L=100$. Number of samples
taken for calculating average values are: $N=1000, 1000, 3000, 4000$
for wire lengths $L=10,20,50,100$ respectively.}
\label{G_rEaverage}
 \end{figure}

Next we discuss the results for the ensemble average conductance and for the
conductance fluctuations. Fig.~\ref{G_rEaverage} shows results for  $\langle
G\rangle$ and $\exp(\langle \ln~G\rangle)$. Conductance quantization disappears
very  quickly, although the presence of a short plateau for the first
subband can be observed for the shorter wires. The boundary roughness  has less
impact for longer wavelengths (smaller energies) and hence the average
conductance tends to decrease with increasing energy.  As the length of the
wire increases there is a rapid decrease of $G$ near the band center. In the
strong localization regime, a broad peak
emerges near the band edge (see Fig.~\ref{G_rEaverage} for $L=100$), which
corresponds to the peak in the localization length\cite{nikolic93}. In this
regime, the average conductance of quasi--1D systems falls off exponentially
with the length \cite{johnston1}:
\begin{equation}
G(E,L)\sim\exp(-2L/\lambda(E))~.
\label{eq:Gexp}
\end{equation}

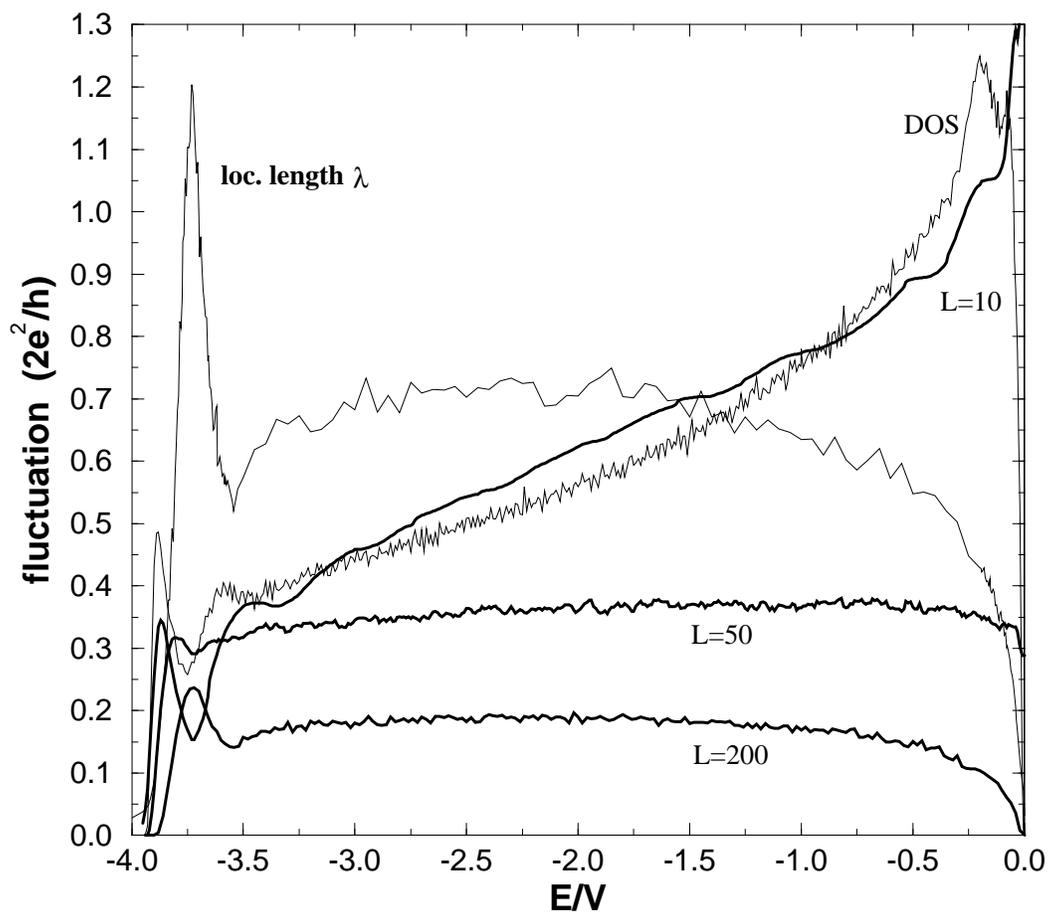
 \begin{figure}
 \caption{Conductance fluctuations which correspond to the case in
Fig.~4 
for the wire lengths: $L=10$, $L=50$ and $L=200$
(three bold lines). Two thin lines are the localization length and the
density of states (both rescaled: $\lambda\rightarrow\lambda/100$ and
${\rm DOS}\rightarrow {\rm DOS}*5$) for quantum wires with boundary
roughness.}
\label{G_rEflc}
 \end{figure}

The conductance fluctuations for the case of the quantum wire with boundary
roughness only, calculated for the examples from Fig.~\ref{G_rEaverage} by
using definition Eq.~(\ref{eq:fluctuation}), are presented in
Fig.~\ref{G_rEflc}.  Three characteristic regimes are shown:
\begin{enumerate}
\item the quasi--ballistic regime --- the wire length is comparable with the
mean free path length, $L\simeq l$ (\eg $L=10$ for  most of the energies and
for the level of disorder assumed  in our samples);
\item the mesoscopic regime where the wire length is $l<L<\lambda$ (\eg
$L=50$);
\item the  strong localization  regime - $L>\lambda$ (\eg $L=200$).
\end{enumerate}
Since the elastic mean free path and the localization length are both functions
of energy,  a particular wire can move between these regimes as the energy is
changed.   The fluctuations for the case $L=10$ in Fig.~\ref{G_rEflc} show
interesting similarities with the density of states for the same type of
quantum wires. There is a peak near the band edge which corresponds to the peak
in the DOS, which is the last remaining feature of the inverse square root
singularities  from the DOS of clean quasi--1D systems (see
Ref.~\cite{nikolic93}). This behaviour can be explained in the following way.
For short wires (\ie quasi--ballistic transport) the important length is the
elastic mean free path $l$. If $l$ increases (as a function of energy) in the
quasi--ballistic regime, then the conductance fluctuations decrease --- as the
scattering in the wire is reduced, and vice versa. Since $l$ is roughly
inversely proportional to the DOS (when the DOS is increased the scattering
rate increases and therefore $l$ decreases), then the  conductance fluctuations
might be expected to mirror the DOS.  As $L$ is increased with respect to $l$
(and still $L<<\lambda$) then the fluctuations will increase towards their
maximal value, which is reached in the mesoscopic regime. The quantum wires of
length $L=50$ in Fig.~\ref{G_rEflc} show a relatively wide region of energy in
which the conductance fluctuations are independent of energy,  with a value
which is close to the universal value for quasi--1D systems ($\rms (G)=0.729
e^2/h$  \cite{lee85}).

For long wires ($L=200$) we have the strong localization regime and $\rms
(G(E))$ follows the curve for the localization length  $\lambda(E)$ (see
Fig.~\ref{G_rEflc}), \ie the average conductance, since these are related
(Eq.~\ref{eq:Gexp}). That $\rms (G)$ and $\lav G\rav$ are proportional can be
understood by using the picture of  `open' and `shut' channels, or `maximal
fluctuations', in a quantum wire in the strong localization regime
\cite{pendry92}.
This terminology is associated with the distribution function $P(\tau)$ for the
eigenvalues $\tau$ ($0\le\tau\le 1$) of ${\bf t}{\bf t}^{\dag}$, where ${\bf
t}$ is the transmission matrix: it has a peak at $\tau=0$, and a tail which
extends to $\tau=1$ \cite{pendry92,mackinnon91}. Each eigenvector of ${\bf
t}_L{\bf t}_L^{\dag}$ defines a conducting  `micro--channel' with the
corresponding conductance $\tau$ in units of  $(2)e^2/h$ \cite{pendry92}. As
the size of the system is increased, the peak at $\tau=0$ strengthens at the
expense of the tail, but the shape of the tail remains the same.

For  long lengths, therefore, most of the micro--channels will have conductance
of order 0 since the bulk of the distribution is around 0.  Only a small
fraction of the channels, with values for $\tau$ of order 1 (roughly
$0.1<\tau\le 1$), will contribute to the conductance.  Such a distribution of
$\tau$, implies that the fluctuations tend to the maximum possible value
consistent with their mean.  However, the moments of the conductance,
$\langle(\Tr {\bf t}{\bf t}^{\dag})^n\rangle$,  for 2D and 3D cases do not show
the same size dependence as the moments of  $\Tr\langle ({\bf t}{\bf
t}^{\dag})^n\rangle$, within the metallic regime.  Universal conductance
fluctuations are affected by the correlations between the $\tau$s rather than
by the distribution of the $\tau$s themselves. UCFs (\ie with a universal value
ca. $e^2/h$) are restricted to disordered systems in the metallic regime where
perturbation theory is applicable. When localization effects are important,
however, the correlations between the micro--channels change and a different
sort of fluctuations is observed \cite{pendry92}. The statistics of a single
micro--channel were found  to be crucial in the strong localization regime.

The average conductance of the disordered quasi--1D system shows an
asymptotically
exponential decrease with the length \cite{johnston1}. This single parameter
dependence suggests that the conductance statistics of quasi--1D samples, with
a length much longer than the localization length,  is dominated by a single
channel. Hence a similar size dependence for the conductance moments should be
expected as for $\Tr\langle ({\bf t}{\bf t}^{\dag})^n\rangle$, since
the distribution of $\tau$s is important, rather than the correlation between
them. This picture of the conductance statistics in the strong localization
regime shows that the conductance fluctuations are proportional to the average
conductance.

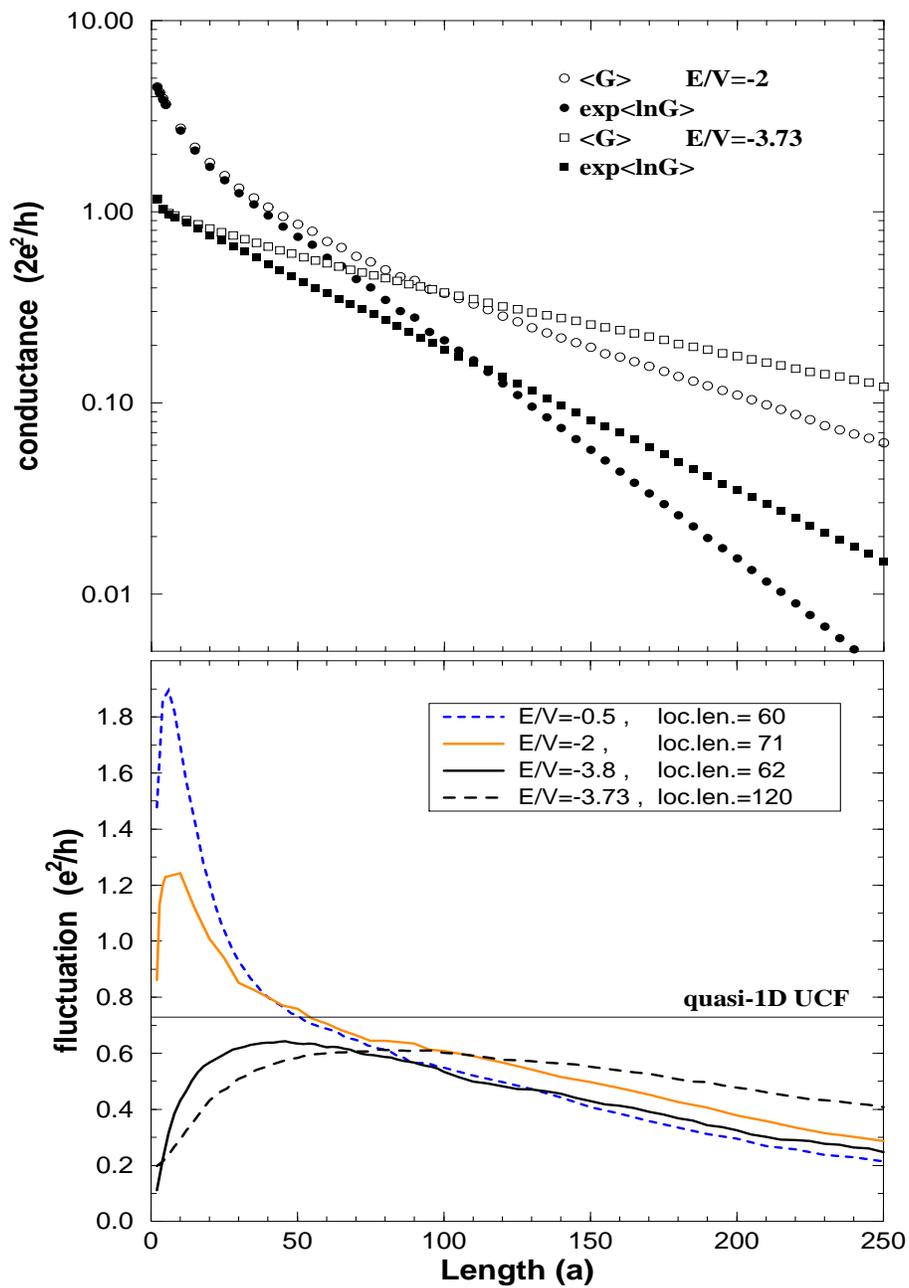
 \begin{figure}
 \caption{Average conductance as a function of wire length for real wire
without islands for energies $E=-2$ and $E=-3.73$ - top figure.
The corresponding conductance fluctuations are
shown in the lower figure as well as fluctuations for energies $E=-0.5$
and $E=-3.8$. Number of samples is $N=4000$.}
\label{G_rLaverage}
 \end{figure}

In Fig.~\ref{G_rLaverage} results are given for the conductance as a function
of wire length, for a fixed energy.  The results for the average conductance
are presented for two values of the energy and two more examples of conductance
fluctuations are added.  The decrease of conductance is more marked in the
quasi--ballistic (near ballistic) regime (short wires), where the difference
between $\langle G\rangle$ and $\exp(\langle\ln G\rangle)$ is negligible.
The difference between the two averages increases as the disorder or length of
the wire is increased (Figures~\ref{G_rEaverage} and \ref{G_rLaverage}). This
shows that the average $\langle G \rangle$ is dominated by a small number of
samples with conductance of order $2e^2/h$.  The divergence of the two curves
indicates that the conductance distribution is transforming to the form which
has a peak for small conductances and a long tail towards larger conductances,
\ie it is transforming to a log--normal distribution \cite{johnston1}.
For wires of  length $L>\lambda$ the decrease of the average conductance
$\exp(\langle\ln G\rangle)$ with the length of the wire is mainly determined by
the localization length. The value for the localization length can be obtained
from Fig.~\ref{G_rLaverage}, where $\langle\ln G\rangle$ is fitted with a
straight line and $\lambda$ is determined from its slope,
\begin{equation}
\lambda = - \frac{\partial \langle\ln G\rangle}{2\partial L} .
\label{eq:lag}
\end{equation}
for wires of length $L>>\lambda$.

The results for the conductance fluctuations are also shown in
Fig.~\ref{G_rLaverage}.  The fluctuations first increase for very short wire
lengths ($L<10$). This is the nearly ballistic regime where the fluctuations
grow as the disorder (or wire length) increases. On the other side, for long
wires, \ie the strong localization regime ($L>\lambda$), the fluctuations
decrease as the wire
becomes longer, due to the overall decrease of the conductance. In this regime
the fluctuations between wires of the same length depend only on the
localization length (see Fig.~\ref{G_rLaverage}).   The fluctuations take
maximal values between these two regimes.
Fluctuations for wires with many subbands ($E=-0.5$ and $E=-2$ in
Fig.~\ref{G_rLaverage}) have a sharp peak for short lengths and then start to
decrease well before the wire has become longer than  $\lambda$. The
fluctuations in the wires with a few subbands ($E=-3.73$ and $E=-3.8$
Fig.~\ref{G_rLaverage}) even appear to be independent of  length for several
ranges of $L$ in the intermediate region.  This could be called the universal
region \cite{ando92,tamura91}, although the constant value of the conductance
fluctuations is not equal to the universal value
obtained for quasi--1D systems in the perturbation theory \cite{lee85}.

\section{The Influence of Islands} \label{conduc-islands}
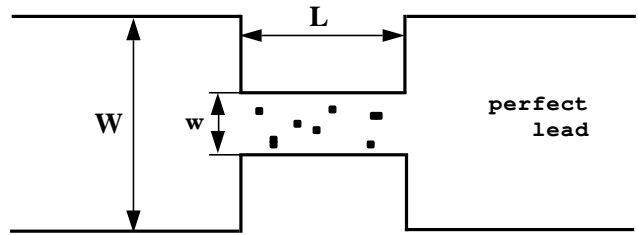
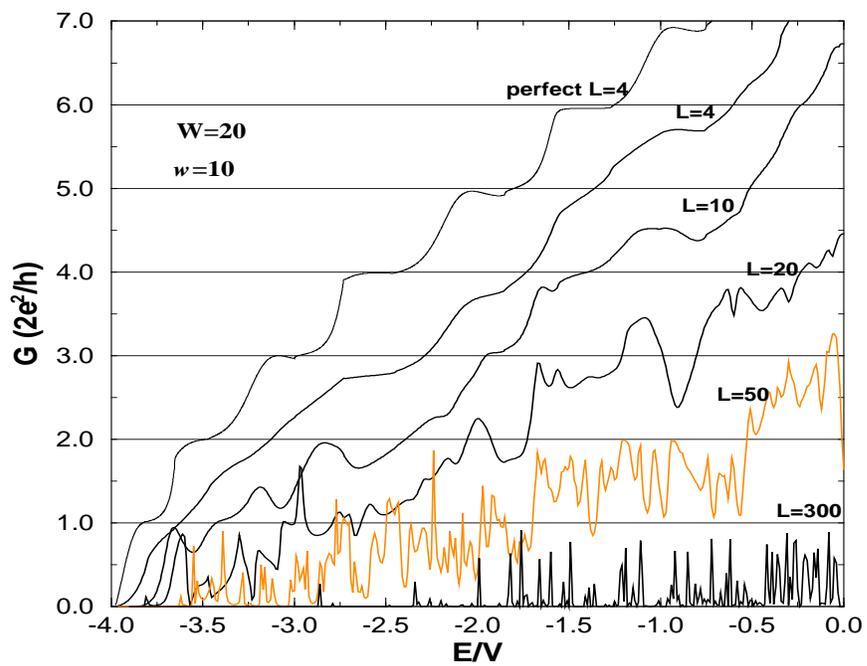
 \begin{figure}
 \caption{$G$ as a function of energy for a single sample of perfect
quantum wire
(width $w=10$) with island disorder, for different wire lengths: $L=4$,
$L=10$, $L=20$, $L=50$ and $L=300$. Island concentration is $p=5\%$.
Also is presented conductance for
perfect constriction of length $L=4$ and width $w=10$.
 The top figure shows the geometry of the system.}
\label{iEone}
 \end{figure}

The effects of the strong scattering centers (\ie islands) in the bulk of the
quantum wire are examined for the case of an otherwise perfect wire with
islands for the system geometry shown in Fig.~\ref{iEone}.
As in the case of rough boundaries we consider the conductance of individual
samples as well as the average conductance as a function of energy or wire
length.

The conductance of a single sample of a quantum wire with islands is shown in
Fig.~\ref{iEone}. Even a small concentration of islands causes an almost
complete suppression of the conductance quantization.  In  Fig.~\ref{iEone} one
can compare the results for a quantum wire of length $L=4$ with island
concentration $p=5\%$ (which means only two islands at random positions) with
the conductance for a perfect wire of length $L=4$.  Island disorder reduces
the conductance in a similar way in each subband.  In the mesoscopic regime,
where $l<L<\lambda$ (examples $L=20$ and $50$ in Fig.~\ref{iEone}), the
conductance fluctuates as a function of energy.  These fluctuations are of the
order $e^2/h$.  This is a quantum interference effect in which the scale of the
sensitivity to changes in the energy depends on the length of the wire.
We estimate that this dependence is of the form $E_c\sim1/L^2$, \ie  similar to
UCF, and differs from that of the fluctuations in the quantum wire with rough
boundaries (Sec.~\ref{conduc-rough}).  $E_c$ is the change of the Fermi energy
needed to modify the relevant phase differences across the sample by about
$2\pi$.  The fluctuation of the conductance between different samples is also
of order $e^2/h$.  In the strong localization regime (\eg case $L=300$ in
Fig.\ref{iEone}) the conductance is reduced to a set of peaks of  maximum
amplitude $2e^2/h$. Each peak corresponds to the occurrence of resonant
tunnelling through the wire.  The localization length generally increases as
energy grows and therefore the peaks get higher towards the center of the band.

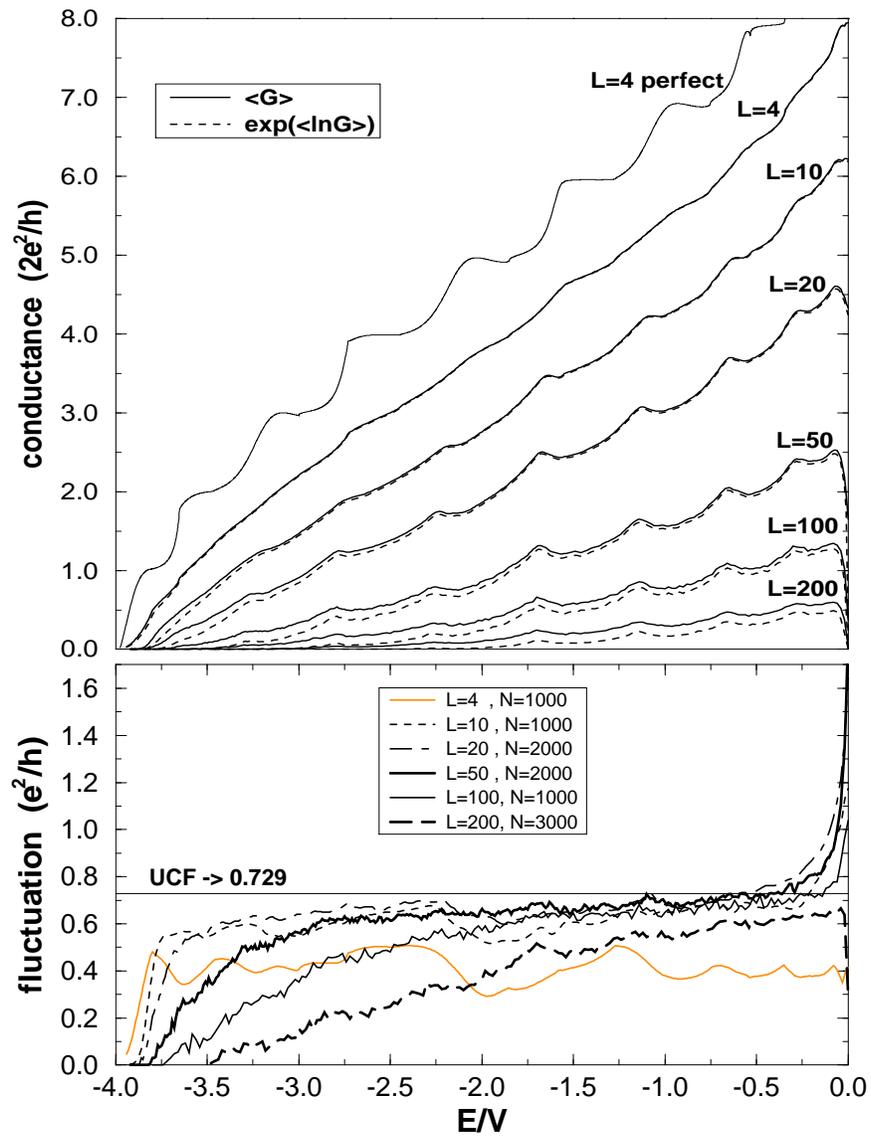
 \begin{figure}
 \caption{Energy dependence of the average conductance and
conductance fluctuations of perfect wire with islands.  Wire lengths are:
$L=4,10,20,50,100,200$. Island concentration is $p=5\%$, $W=20$ and wire
width is $w=10$. Number of samples
taken for calculating average values is N. The universal value for UCF of
quasi-1D metallic wires is marked by a horizontal line.}
\label{G_iEaverage}
 \end{figure}

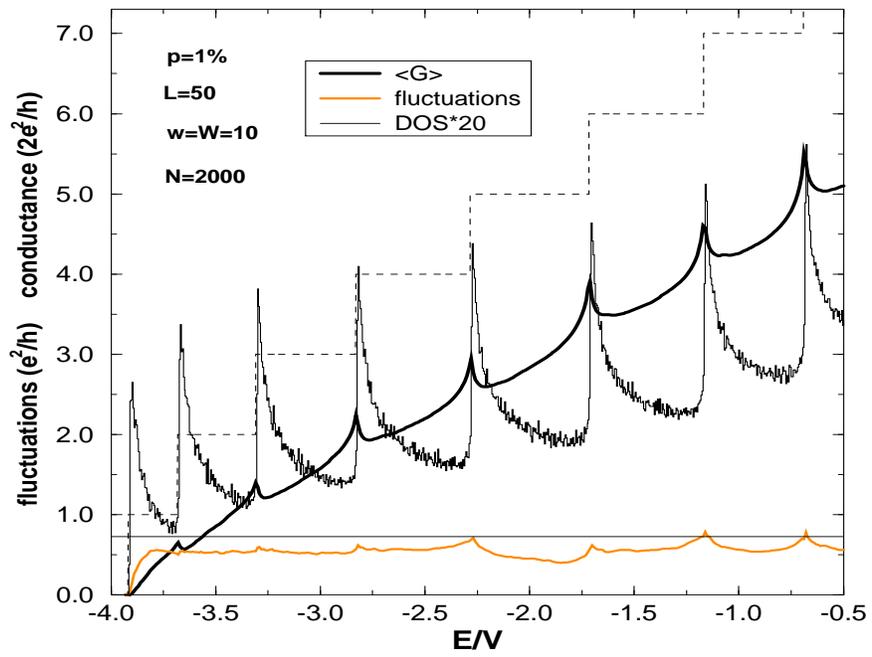
 \begin{figure}
 \caption{$\langle G(E)\rangle$ and $\rms (G(E))$ for a perfect wire
with island concentration $p=1\%$. Steps represent the conductance for the
case of perfect wire. (Note the different units for $\lav G\rav$
and fluctuations.) The histogram of the (rescaled) DOS for this wire is also
shown.}
\label{i1Eaverage}
 \end{figure}

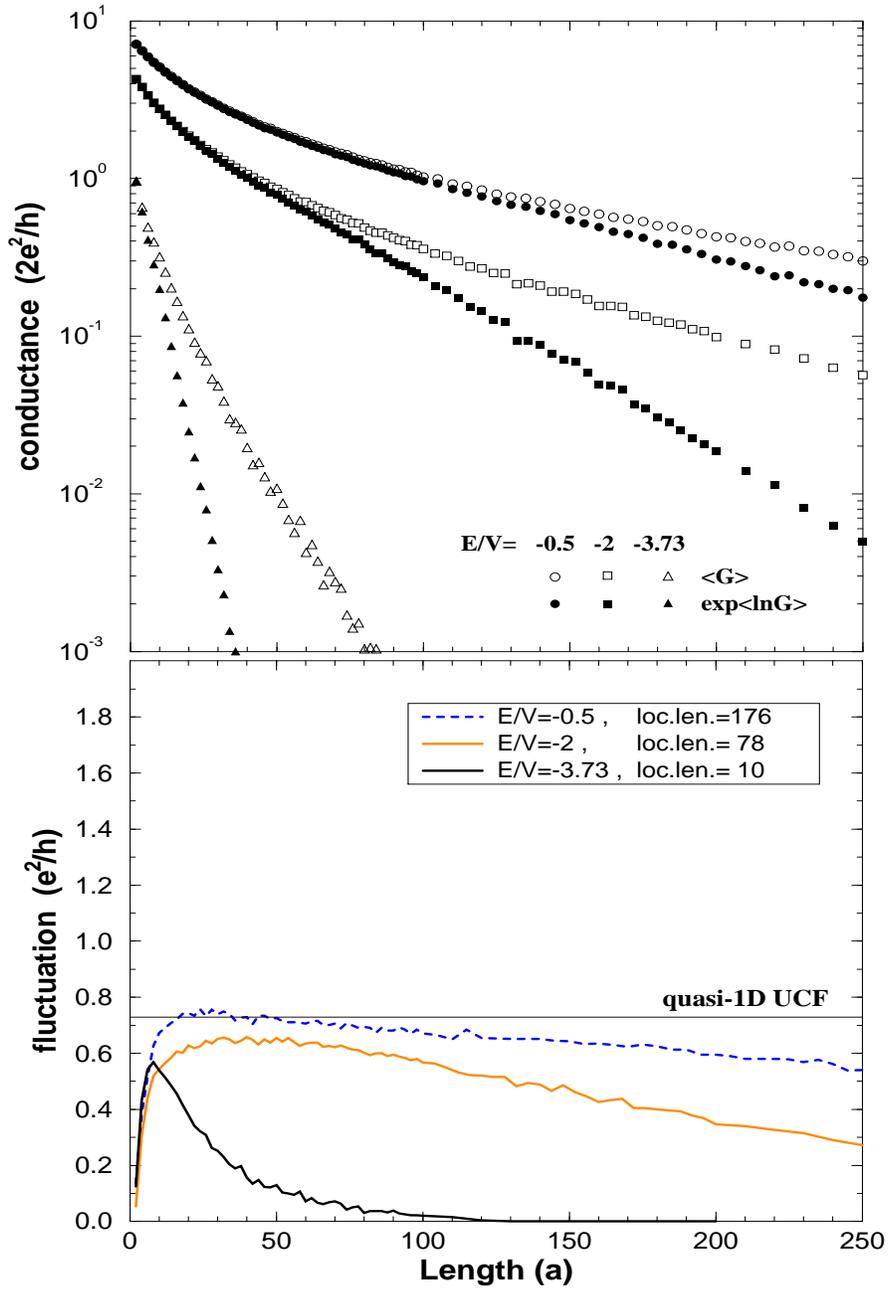
 \begin{figure}
 \caption{Average conductance as a function of wire length for perfect wire
with islands (concentration $p=5\%$) for energies $E=-0.5,-2$ and $-3.73$ -
upper figure. The corresponding conductance fluctuations are shown in the
lower figure. Number of samples is $N=4000$, $W=20$ and $w=10$.}
\label{G_iLaverage}
 \end{figure}

The average conductance for the case of bulk (island) disorder is shown in
Fig.~\ref{G_iEaverage} and Fig.~\ref{G_iLaverage}. This type of disorder has a
similar effect on each subband, unlike the case of edge roughness where higher
subbands are more affected then lower ones. The average conductance as a
function of energy exhibits local maxima near the energies of the subband edges
of the perfect wire (Figures~\ref{G_iEaverage} and \ref{i1Eaverage}). This
becomes more obvious for the smaller concentrations of islands (\eg the island
concentrations are $p=5\%$ in the case presented in Fig.~\ref{G_iEaverage} and
$p=1\%$ in Fig.~\ref{i1Eaverage}).

The behavior of the average conductance of a disordered quantum wire modelled
by the (Anderson) Hamiltonian with a uniform distribution for the site energies
of the wire \cite{masek89} is, to some extent, similar. Although the general
appearance of the curves differs (for the strong--scattering regime it looks
like a line with peaks, whereas for the Anderson model it looks like a line
with dips), in both cases the DOS is connected to $\lav G(E)\rav$ in the same
way.  Electron scattering is proportional to the number of available states
into which an electron can be scattered, \ie to the DOS. Therefore the electron
mobility and conductance should decrease when the DOS increases and vice versa,
see Fig.~\ref{i1Eaverage}.

The average conductance decreases exponentially with the length of the wire,
see Fig.~\ref{G_iLaverage}, which is the expected behavior for the localized
states. The slope of the line  $\exp(\lav \ln (G)\rav)$ for long wires
determines the localization length for any energy, as defined by the
Eq.~(\ref{eq:lag}).

The conductance fluctuations first increase in the quasi--ballistic regime
(Fig.~\ref{G_iLaverage}), go through a  maximum in the region $l<L<\lambda$ and
then decrease as the length of the wire increases. The decrease is slower for
energies with longer localization lengths. A short region of lengths,  where
fluctuations are almost independent of the wire length, can be observed for the
energies $E=-0.5$ and $E=-2$. The level of the fluctuations in this universal
region depends on energy and it is not, in general, equal to UCF value for
quasi--1D metallic systems. However, by chance we have found a case (that is
$E=-0.5$ in Fig.~\ref{G_iLaverage}) where $\rms (G) \approx 0.73$.  The
conductance fluctuations (Fig.~\ref{G_iEaverage}) increase with the energy and
tend to a sort of asymptotical value. This value is close to the UCF constant
for wires with lengths inside the region $l<L<\lambda$, and decreases as the
wire lengths move out of this region.

\section{Real wires with islands}  \label{conduc-real}
So far we have examined separately the effects of boundary roughness and
islands on the quantum wire conductance. Real wires with islands have both
types of disorder. The geometry of the system used for the calculation of the
conductance is the same as in the case of real wires without islands, shown at
the top of Fig.~\ref{rEone}.

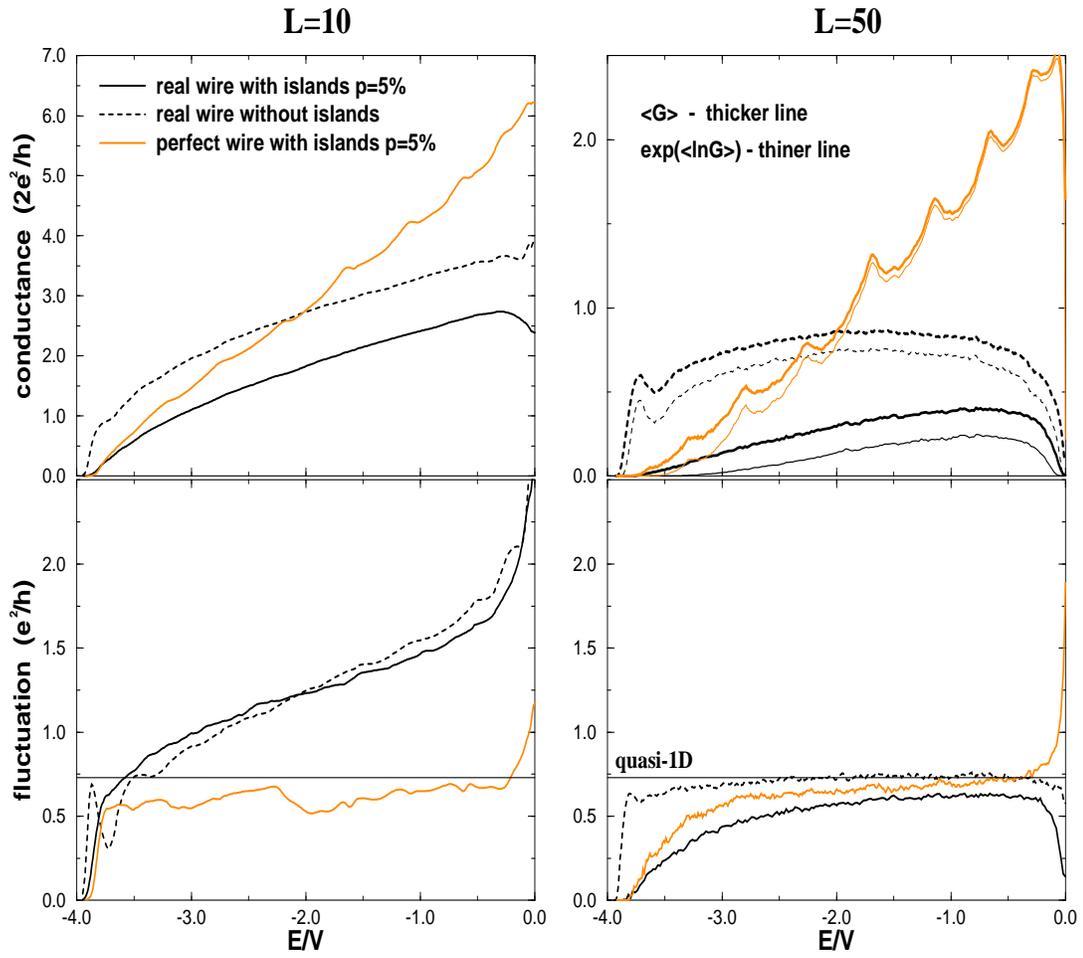
 \begin{figure}
 \caption{The average conductance and the conductance fluctuations as
function of energy for real wire with islands, p=5\% (black full line), for
wire lengths $L=10$ and $L=50$. These results can be compared with the
results for the case of real wire without islands (\ie wire with rough edges),
$W=20$, $\lav w\rav=10$, and perfect wire with islands of concentration
$p=5\%$ and $W=20$, $w=10$. For wire length $L=10$, $\lav G\rav$
and $\exp(\lav\ln G\rav)$ are virtually the same, so only $\lav G\rav$ is
shown, whereas for $L=50$ both averages are shown. Number of samples are:
$N=1000$ $(L=10)$ and $N=3000$ $(L=50)$.}
\label{VI_L10a}
 \end{figure}

Some results  for the average conductance and fluctuations as functions of
energy are shown in Fig.~\ref{VI_L10a} for two lengths of wire, $L=10$ and
$L=50$. The average conductance for both regimes shows no features, just a
monotonically rising curve which bends and falls off near the band centre. The
presence of both types of disorder causes a further decrease of the conductance
when compared with a wire with only one type of disorder. One can say that the
influence of islands is dominant for energies near the band edge, whereas the
influence of edge roughness becomes dominant towards the band centre. Any
feature of the conductance quantization for a single sample is quickly
destroyed by the presence of both types of disorder. The conductance
fluctuations for nearly ballistic samples ($L=10$) have similar values to the
fluctuations for the case of a real wire without islands, except for energies
near the band edge. However, as the wire length increases the conductance
fluctuations for a real wire with islands decreases faster then for the case
without islands. This is because strong localization sets in sooner in the
presence of islands. It was found in  Ref.~\cite{nikolic93} that the
localization length of real wires with islands is usually about half of the
localization length of real wires without islands (case of $p=5\%$ island
concentration).

The conductance fluctuations as a function of length behave like the
conductance fluctuations for the case of a perfect wire with islands for
energies
near the band edge, but for higher energies they are similar to the case of
real wires without islands.

\section{Discussion and Conclusion} \label{conduc-concl}
 The characteristic coherent transport regimes (quasi--ballistic, mesoscopic
--- $l<L<\lambda$ and strong localization) are each affected by disorder in a
similar way. For boundary roughness the influence is weak near the band edge,
and increases as the energy increases. However, the influence of the island
disorder is strongest on the highest propagating modes and does not depend on
the mode (subband) number. Since the total conductance increases with the
number of propagating modes, then, in general, the average conductance
increases with the energy. But note that the average conductance, as a function
of energy, always drops when a new subband opens due to the enhanced
intersubband scattering. This was not observed for the case of boundary
roughness only.  For the quasi--ballistic regime the conductance quantization
deteriorates very rapidly as the number of scattering events increases.  The
average conductance decays exponentially as a function of wire length (for
$L>\lambda$), for any kind of disorder, which is considered as an additional
confirmation of the exponential localization of electron states. Anderson
localization is very effective in reducing the carrier mobility in narrow
quantum wires. This effect acts strongly against the predicted high mobility
for quantum wires \cite{sakaki80}.

The conductance fluctuations of narrow quantum wires depend, in general, on the
length of the wire, and are therefore not universal conductance fluctuations
(UCF). However, the case with rough edges can show a universal region, but only
for energies in the first subband. The value for $\rms (G)$ is not far from the
UCF value for metallic quasi--1D systems ($0.729e^2/h$) and depends on the
energy.  An increase in the localization length extends this region of
constant fluctuations.  Ando and Tamura \cite{ando92} have predicted that for
wider wires than we have, a much broader region of universal conductance
fluctuations will appear. We find, on the contrary, that for island disorder
short universal regions can exist only for higher energies and the actual value
for $\rms (G)$ approaches the UCF as the energy increases. The universal
region, if it exists, appears for  wires of length $l<L<\lambda$, and the
fluctuations reach a maximum in this region.

Changing  the cross--section of the leads makes no qualitative difference to
the conductance of a system consisting of a disordered wire attached to two
perfect leads \cite{PhDThesis}. Some small quantitative changes are observed
only when $W\sim w$ (which will be reported elsewhere). In any case the
conductance becomes independent of $W$ when $W>>w$. This should be expected,
since for large values of $W$, transverse modes in the leads are densely
distributed, and there are many of them contributing to the total conductance
\cite{szafer89}.

\begin{references}

\bibitem{petroff84} P. M. Petroff, A. C. Gossard and W. Weigmann, \apl{45},
 620 (1984).

\bibitem{vanwees88} B. J. van Wees, H. van Houten, C. W. Beenakker, J. G.
 Williamson, L. P. Kouwenhoven, D. van der Marel and C. T. Foxton,
 \prl{60}, 848 (1988); D. A. Wharam, T. J. Thornton, R. Newbury, M. Pepper,
H. Ahmed, J. E. F. Frost, D. G. Hasko, D. C. Peacock, D. A. Ritchie and G.
A. C. Jones, \jpcm{21}, L209 (1988).

\bibitem{glazman88} L. I. Glazman, G. B. Lesovik, D. E. Khmel'nitskii and
 R. I. Shekhter, {\it Pis'ma Zh. Eksp. Teor. Fiz.} {\bf 48}, 218 (1988);
 L. I. Glazman and M. Jonson, \prb{41}, 10686 (1990).

\bibitem{kirczenow88} G. Kirczenow, {\it Solid State Commmun.} {\bf 68},
 715 (1988); \prb{39}, 10452  (1989); A. Szafer and A. D. Stone, \prl{62},
 300 (1989); Song He and S. Das Sarma, \prb{48}, 4629 (1993).
\bibitem{PhDThesis} K. Nikoli\'c, PhD Thesis, Imperial College, London
 (1993).

\bibitem{timp92} G. Timp, {\it "Physics of Nanostructures"} ed. J. H.
 Davies and A. R. Long, Proc. of 38th SUSSP, IOP: Bristol and Philadelphia
 (1992).

\bibitem{nixon91} J. A. Nixon, J. H. Davies and H. U. Baranger, \prb{43},
 12638 (1991).
\bibitem{laughton91} M. J. Laughton, J. R. Barker, J. A. Nixon and J. H.
 Davies, \prb{44}, 1150 (1991).

\bibitem{sakaki80} H. Sakaki, {\it Jpn.\ J.\ Appl.\ Phys.\ }{\bf 19}, 94
 (1980).

\bibitem{tamura91} H. Tamura and T. Ando, \prb{44}, 1792 (1991).
\bibitem{ando92} T. Ando and H. Tamura, \prb{46}, 2332 (1992).

\bibitem{taylor91} J. P. G. Taylor, K. J. Hugill, D. D. Vvedensky and A.
  MacKinnon, \prl{67} 2359 (1991).

\bibitem{stone92} A. D. Stone, in
{\it "Physics of Nanostructures"} ed. J. H. Davies and A. R. Long, Proc. of
38th SUSSP, IOP: Bristol and Philadelphia (1992).

\bibitem{hugill} K. J. Hugill, S. Clarke, D. D. Vvedensky and B. A. Joyce,
 {\it J.\ Appl.\ Phys.} {\bf 66}, 3415 (1989).
\bibitem{fukui87} T. Fukui and H. Saito, \apl{50}, 824 (1987).
\bibitem{miller92} M. S. Miller, H. Weman, C. E. Pryor, M. Krichnamurthy,
 P. M. Petroff, H. Kroemer and J. L. Merz, \prl{68}, 3464 (1992).
\bibitem{notzel91} R. N\"{o}tzel, N. N. Ledentsov, L. D\"{a}weritz, M.
 Hohenstein and K. Ploog, \prl{67}, 3812 (1991).

\bibitem{example} These widths of about $10$nm yield a subband separation of
$\sim 100$ meV for the lateral confinement potential of several hundred meV.

\bibitem{joyce} B. A. Joyce, J. H. Neave, J. Zhang, D. D. Vvedensky, S.
 Clarke, K. J. Hugill, T. Shitara and A. K. Myers-Beaghton,
 {\it Semicond. Sci. Technol.} {\bf 5}, 1147 (1990).

\bibitem{ando91} T. Ando,\prb{44}, 8017 (1991).
\bibitem{baranger89} H. U. Baranger and A. D. Stone, \prb{40}, 8169
 (1989).

\bibitem{mackinnonZ85} A. MacKinnon, \zpb{59}, 385 (1985).
\bibitem{mackinnonprl81} A. MacKinnon and B. Kramer, \prl{47}, 1546 (1981).
\bibitem{soukoulis82}  C. M. Soukoulis, I. Webman, G. S. Grest and E. N.
  Economou', \prb{26}, 1838 (1982); A. D. Zdetsis, C. M.
  Soukoulis, E. N. Economou and G. S. Grest, \prb{32}, 7811 (1985).

\bibitem{landauer57} R. Landauer, \ibmj{1}, 223 (1957); {\it Philos.
 Mag.\ }{\bf 21}, 863 (1970).
\bibitem{fisher81} D. S. Fisher and P. A. Lee, \prb{23}, 6851 (1981).

\bibitem{lee85} P. A. Lee and A. Douglas Stone, \prl{55}, 1622 (1985);
 P. A. Lee, A. D. Stone and H. Fukuyama, \prb{35}, 1039 (1987).
\bibitem{takagaki92} Y. Takagaki and D. K. Ferry, \jpcm{4}, 10421 (1992);
 \prb{46}, 15218 (1992).

\bibitem{bryant91} G. W. Bryant, \prb{44}, 12837 (1991).
\bibitem{johnston1} R. Johnston and H. Kunz, \jpc{16}, 3895 (1983).
\bibitem{nikolic93} K. Nikoli\'c and A. MacKinnon, \prb{47}, 6555 (1993).

\bibitem{pendry92} J. B. Pendry, A. MacKinnon and P. J. Roberts, \prs{A437}
 (1992).
\bibitem{mackinnon91} A. MacKinnon, in "{\it Quantum Coherence in
 Mesoscopic Systems}", ed. B. Kramer, Plenum Press, New York 1991,
 pp. 415-427.

\bibitem{masek89} J. Ma\v sek and B. Kramer, \jpcm{1}, 6395 (1989).

\bibitem{szafer89} A. Szafer and A. D. Stone, \prl{62}, 300 (1989).

\bibitem{Pendry} J. b. Pendry \jpc{20},  {733} 1987
\end{references}
\end{document}